\documentstyle [aps,prl,psfig,preprint,epsfig]{revtex}

\tightenlines

\begin{document}
\widetext \title{ Matrix element effects in angle-resolved
  photoemission from Bi2212: Energy and polarization dependencies,
  final state spectrum, spectral signatures of specific transitions
  and related issues } 

\author{ M. Lindroos$^{1,2}$, S. Sahrakorpi $^{1,2}$ and A. Bansil$^2$
  } 

\address{$^1$ Tampere University of Technology, P.O. Box 692,
  FIN-33101 Tampere, Finland}

\address{$^2$ Physics Department, Northeastern University, Boston,
  Massachusetts 02115}

\maketitle 
\widetext

\begin{abstract}
  
  We have carried out extensive simulations of the angle-resolved
  photoemission (ARPES) intensity in Bi2212 within the one-step and
  three-step type models using a first-principles band theory
  framework. The focus is on understanding the behavior of emissions
  from the antibonding and bonding bands arising from the $CuO_2$
  bilayers around the $\overline{M}(\pi,0)$ symmetry point. The
  specific issues addressed include: Dependencies of the
  photointensity on the energy and polarization of the incident light;
  character of the initial and final states involved as well as the
  spectrum of the relevant final states; and, changes in the spectral
  intensity as a function of the perpendicular component, $k_\perp$,
  of the momentum of the photoelectron. Considerable insight into the
  nature of individual transitions is adduced by examining the
  momentum matrix element for bulk transitions within the solid and by
  further decomposing this matrix element into contributions from
  various atomic sites and angular momentum channels.  These results
  indicate that, via remarkable interference effects, the ARPES matrix
  element can in particular cases help zoom in on the properties of
  electrons excited from specific sites and/or angular momentum
  channels even in a complex material.

\end{abstract}


\narrowtext

\section{Introduction}

It is well recognized that the photoemission process involves an
interplay between the bulk and surface phenomena as the excited
electron must be ejected from the surface in order to reach the
detector. The intensity measured in an ARPES (angle-resolved
photoemission spectroscopy) experiment, therefore, fundamentally
involves the properties of the relevant initial and final states in
the presence of the surface.  As a result, the ARPES intensity can
differ greatly from the spectral density of the initial state involved
in the excitation. An understanding of this difference $-$ the
so-called 'matrix element effect', is essential for a satisfactory
interpretation of the ARPES spectra, and this is particularly so in
complex materials such as the high-Tc's\cite{matri,mlcape,maria}.

Bearing these considerations in mind, we have carried out extensive
simulations of the ARPES spectra in $Bi_2Sr_2CaCu_2O_8$ (Bi2212,
BISCO) within the first-principles band theory framework. The pristine
phase with the body centered tetragonal lattice structure is assumed;
the computationally more demanding case of the orthorhombic and
modulated lattices is hoped to be taken up in the future. We focus on
the excitation of states in the vicinity of the Fermi energy ($E_F$)
around the $\overline{M}(\pi,0)$ symmetry point which have been the
subject of intense scrutiny and controversy in the high-Tc literature.
For example, the issue of bilayer splitting$-$ expected to be
relatively large within the conventional LDA-based band theory picture
at $\overline{M}$ in BISCO, has been debated for some time, even
though a consensus in favor of the existence of a bilayer splitting
appears to be emerging in the recent months\cite{zx2,zx3,des2}.  The
fundamental question has been whether or not strong correlations
localize electrons in the $CuO_2$ layers, yielding a 2D electron gas
and a zero bilayer splitting with significant implications for the
mechanism of superconductivity in the cuprates.\cite{gene1} This study
gives insight into these and related aspects of the ground state
electronic structure by delineating the connection between the ARPES
spectra and the character of the underlying initial and final states.

For this purpose, we have developed the one-step methodology
\cite{pendry2} for modeling the photoemission process so that
arbitrarily complex systems with many basis atoms can be treated
\cite{matri,mlcape,stanf,ncco,cucis}. The previous one-step work has
been limited to a maximum of two atoms per unit cell \cite{larsson}
which is not adequate for the cuprates; BISCO, for example, involves
30 atoms per conventional unit cell even for the pristine lattice. In
the one-step approach, the interaction of light with the solid is
treated as a single quantum mechanical event and the artifical
distinctions between the processes of excitation, transport and
ejection of the photoelectron invoked in the earlier three-step models
\cite{cardona78,kunz79,inglesfield92,lindroos82,ni100} are not made.
The one-step scheme is thus inherently more satisfactory than the
three-step model since the surface is incorporated in the latter in an
{\it ad hoc} manner. On the other hand, it often tends to be difficult
to identify the behavior of individual transitions within the one-step
methodology where the initial and final states must be damped, causing
these states to broaden and overlap, blurring distinctions between
various transitions. For this reason, we have also developed the
three-step model for complex lattices\cite{ni100}.  Useful insight is
adduced here into different contributions to the ARPES spectrum of
BISCO by examining the momentum matrix element for bulk transitions in
the solid in the spirit of the three-step approach.

The specifics of some of the main issues addressed in this article and
highlights of the new results presented are as follows. The
dependencies of the ARPES intensity on the energy and polarization of
the incident light are clarified for the excitation of states near the
$\overline{M}$ point. The response of the antibonding and bonding
combinations of states from the $CuO_2$ bilayers is found to display a
strong energy and polarization dependence. Moreover, the behavior of
the antibonding and bonding states differs significantly so that by a
judicious choice of these experimental parameters even closely placed
bilayer bands could be distinguished. Also, photon energies where the
intrinsic cross-section for exciting certain states of the pristine
lattice is large will be best suited for studying these states; in
contrast, when this cross-section is relatively small, the effects of
deviations from the tetragonal symmetry of the system via various
distortions and modulations will become more prominent. Further, we
show how some systematics of the polarization dependencies can be
understood in the cuprates via straightforward arguments involving the
symmetries of the initial and final states.

Turning to questions related to individual transitions, we discuss the
nature of the momentum matrix element for bulk excitations within a
three-step type model in terms of contributions from various basis
atoms and different angular momentum channels. The character of the
bonding and antibonding initial states as well as that of the relevant
final states at $\overline{M}$ is clarified. An analysis of these
results reveals dramatically how different contributions to the
momentum matrix element can interfere constructively or destructively
to enhance or supress the weights of photoelectrons excited from
particular atomic sites and/or specific angular momentum channels.
This remarkable result which, to our knowledge, has not been
recognized previously in the literature, hints that the ARPES matrix
element may allow us to zoom in on the properties of very specific
transitions even in complex unit cell materials.

Concerning relevant literature, Refs.
\onlinecite{matri,mlcape,stanf,ncco,santa,abri} discuss some of our
earlier work on BISCO, $Nd_{2-x}Ce_xCuO_4$ (NCCO), $YBa_2Cu_3O_7$
(YBCO) and $YBa_2Cu_4O_8$(Y124).  In particular, Ref.
\onlinecite{matri} emphasizes the importance of matrix element effects
in ARPES spectra, while Ref. \onlinecite{maria} undertakes an
extensive comparison between the one-step predictions over a wide
region in the momentum space when the initial state energy is held
fixed at the $E_F$ with the corresponding experimental ARPES results
giving the Fermi surface map in BISCO. There is of course a large body
of available work on the photoemission theory and Refs.
\onlinecite{photogen,mahan,caroli,durham,braun,grass} would provide a
brief selection. On the experimental side as well the ARPES studies of
BISCO are far too numerous to be cited with any completeness, and we
refer only to a few representative papers.
\cite{zx2,zx3,des2,camp2,gold1,takahashi,aebi,camp1,%
  bianconi,shen,valla,lindau,onellion}

An outline of this article as is follows. The introductory remarks are
followed in Section II by a consideration of formal matters; a few
equations for the one-step model are presented (Subsection IIA), and
the problem of computing the momentum matrix element in a general
lattice is formulated (Subsection IIB). Section III delineates the
relevant features of the lattice geometry (Subsection IIIA) and the
band structure of BISCO (Subsection IIIB). The results of Sections II
and III serve as a source of frequent reference in the subsequent
discussion. Section IV is divided into several subsections to go over
various salient features of the ARPES spectra from $\overline{M}$. The
aspects addressed concern: Photon energy dependence (Subsection IVA);
Effect of final state width (Subsection IVB); Polarization dependence
(Subsection IVC); and, the final state band structure and momentum
matrix element (Subsection IVD).  Section V takes up the question of
the nature of the transition matrix element and the origin of specific
spectral features in terms of the characters of the initial and final
states and is divided into three parts dealing with: Contributions of
different atomic sites (Subsection VA); contributions of various
angular momentum channels (Subsection VB); and, comments on the
$k_\perp$ dependence of the spectral intensities (Subsection VC).
Finally, Section VI summarizes our results and makes a few concluding
remarks.

\section{Theoretical considerations}

\subsection{General aspects}

We begin by recalling that the ARPES intensity in the one-step model
can be expressed as\cite{caroli,pendry2}
\begin{equation}
{I = -{1\over{\pi}} Im  <{\bf k}_{\parallel}|G_2^+\ \Delta \ G_1^+
\Delta^\dagger \ G_2^-|{\bf k}_{\parallel}>}
\label{y1}
\end{equation}
where the ket $|{\bf k}_{\parallel}>$ denotes a free electron state
with momentum, ${\bf k}_{\parallel}$, parallel to the crystal surface,
$G_2$ ($G_1$) is the retarded (+) or advanced (-) one-electron Green
function for the final (initial) state and
\begin{equation}
\Delta=e\hbar/2mc ({\bf p}\cdot{\bf A}+{\bf A}\cdot{\bf p}) 
\label{y2}
\end{equation}
is the interaction Hamiltonian in terms of the vector potential ${\bf
  A}$ of the incident photon field and the electron momentum operator
${\bf p}$.

For our purposes, we rewrite Eq. \ref{y1} by using the representation 
\begin{equation} 
-{1 \over \pi} Im G_1^+ = \sum_i |i> B_{ii}<i|
\label{y3}
\end{equation}
of the Green function $G_1$ as a sum over the spectral functions $
B_{ii}$ for various initial states $i$ which yields
\begin{eqnarray} 
I& = & <{\bf k}_{\parallel}|G_2^+ \Delta \sum_i |i> B_{ii}<i|\Delta^\dagger  
G_2^-|{\bf k}_{\parallel}>\\
& =  &\sum_i B_{ii} <{\bf k}_{\parallel}|G_2^+\Delta|i><i|\Delta^\dagger \ 
G_2^-|{\bf k}_{\parallel}> \\
& = & \sum_i B_{ii} |<f|\Delta|i>|^2
\label{y4}
\end{eqnarray}
Here we have replaced $G_2^{-}|{\bf k}_{\parallel}>$ by the final
state wavefunction $|f>$.  Substituting form for $\Delta$ in Eq.
\ref{y2}, and neglecting the term proportional to ${\bf \nabla} \cdot
{\bf A}$ by assuming that the vector potential A varies smoothly, i.e.
wavelength of light is much larger than the unit cell
dimensions\cite{footmatti}, we obtain
\begin{equation} 
I =  {e\hbar}/{mc}\sum_i B_{ii} |{\bf A} \cdot <f|{\bf p}|i>|^2
\label{y5}
\end{equation}
Notably, the summation over $i$ not only encompasses different initial
state bands, but also involves an integration over ${\bf k}_{\perp}$.
A computationally practical formula follows by invoking the identity
\begin{equation}[H,p] \equiv Hp-pH= -i\hbar\nabla V
\label{y6}
\end{equation}
which is valid when the potential term is ${\bf p}$ independent.  The
matrix element in Eq. \ref{y5} is easily recast using Eq. \ref{y6} as
\begin{equation}{\bf A} \cdot <f|{\bf p}|i> =
\frac{-i\hbar}{E_i-E_f}{\bf A} \cdot<f|{\bf \nabla} V|i>
\label{y7}
\end{equation}
where $E_i$ and $E_f$ are the initial and final state energies,
respectively.

\subsection{Momentum matrix element}

The nature of the photointensity based on the one-step Eq. \ref{y1}
can be delineated in terms of the momentum matrix element $<f|{\bf
  p}|i>$ of Eq. \ref{y5}. However, the wavefunctions $|i>$ and $|f>$
occuring in Eqs. \ref{y3} - \ref{y7} are complicated quantities which
include multiple scattering effects in the presence of the surface and
damped initial and final state propagators.  Although the full
one-step calculation should always be kept in mind, insight into the
results may be obtained by simplifying the situation and replacing
$|i>$ and $|f>$ by the corresponding Bloch wavefunctions
$\tilde{\psi}$ in the {\it bulk} crystal and the resulting matrix
element $<\tilde{\psi}_f|{\bf p}|\tilde{\psi}_i>$ for optical
transitions.  It should be emphasized that the optical matrix element
must in general be supplemented with other information specific to the
photoemission experiment (e.g. symmetry of the final state) before
connection with a particular measurement can be made. This issue will
be addressed as needed in the remainder of this article.

The formalism for evaluating the optical matrix element for a general
lattice within the KKR scheme is given in Refs.
\onlinecite{ni100,seppo}; an outline of aspects relevant for the
purposes of this article is as follows.  The starting point is the KKR
wavefunction for a general lattice which can be expressed within the
unit cell as \cite{segall,peter}
\begin{equation}
\tilde{\psi}({\mathbf r}) =  
\sum_{L,\beta} \> i^l C_L^\beta R_l^\beta(r) Y_L(\Omega)
\label{y8}
\end{equation}
where $L \equiv (l,m)$ is a composite angular momentum index and
$\beta$ denotes different basis sites.  $C_L^\beta$ are expansion
coefficients and $R_l^\beta(r)$ is the radial part of the Bloch
wavefunction on site $\beta$ in angular momentum channel $L$.
$Y_L(\Omega)$ are real spherical harmonics.  Using expressions of Eq.
\ref{y8} for the initial and final states yields
\begin{equation}
 <\tilde{\psi}_f|{\bf p}|\tilde{\psi}_i> = \sum_{\alpha ,\beta} \sum_{L,L'} 
{\mathbf {\hat{e}}}_\alpha i^{l-l'-1} {C_{L'}^\beta}^* C_L^\beta  
B_{l,l'}^\beta {\cal G}_{L,L'}^\alpha
 \label{y9}
\end{equation}
where the primed indices refer to the final state and ${\mathbf
  {\hat{e}}}_\alpha$ is a unit vector along the direction $\alpha $.
$B_{l,l'}^\beta $ involves an integral over the radial part of the
initial and final state wavefunction and their derivatives and is
given by
\begin{equation}
B_{l,l'}^{(\beta)} = \left\{
\begin{array}{ll} 
{\displaystyle \int}_0^{r_\beta} 
\left( P_{l'}^\star P_l' - 
\displaystyle { l + 1 \over r } 
P_{l'}^\star P_l \right)  dr \\\hspace*{1cm}
+ \displaystyle {1 \over E_i - E_{f}}   
\\ \hspace*{1cm}
\times \left\{ P_{l'}^\star P_l 
\left[ \displaystyle {(l+1)^2 \over r^2 } - E_i \right] \right.
 \\ \hspace*{1cm}
+
\displaystyle { l + 1   \over r } 
\left[ { P_{l'}' }^\star P_l - 
P_{l'}^\star P_l' 
\right]
  \\ \hspace*{1cm} 
\left. - { P_{l'}' }^\star P_l' 
\right\}_{r=r_\beta} & l' = l + 1 
\\        
{\displaystyle \int}_0^{r_\beta} 
\left( P_{l'}^\star P_l' 
+ \displaystyle { l \over r } P_{l'}^\star P_l \right)  dr
\\ \hspace*{1cm}+ 
\displaystyle { 1 \over E_i - E_{f} }   
\\ \hspace*{1cm}
\times \left\{ 
P_{l'}^\star P_l 
\left[ \displaystyle { l^2 \over r^2 } - E_i \right] \right.
 \\ \hspace*{1cm}
+ \displaystyle { l \over r } 
\left[ { P_{l'}' }^\star P_l - 
P_{l'}^\star P_l' \right]
\\ \hspace*{1cm}
\left. - { P_{l'}' }^\star P_l' 
\right\}_{r=r_\beta} & l' = l - 1 
\end{array} ~, 
\right.
\label{y10}
\end{equation}
Here, $P_l(r) \equiv rR_l^{\beta}(r)$, 
$r_\beta$ is the radius of the $\beta^{th}$ muffin-tin sphere, 
$E_i$ and $E_f$ have been defined in connection with Eq. \ref{y7} above, and 
\begin{equation}
{\cal G}_{L,L'}^\alpha = \left\{
\begin{array}{lr}
\sqrt{ 4\pi \over 3 } 
\displaystyle \int Y_{L'}(\Omega) Y_{1,\alpha}(\Omega) 
Y_L(\Omega) d\Omega \>\>\>
& l' = l \pm 1 \\
0 & l' \neq l \pm 1 
\end{array} ~,
\right. 
\label{y11}
\end{equation}
${\cal G}_{L,L'}^\alpha$ are seen to contain angular momentum
selection rules for dipole transitions. $Y_{1,\alpha}$ for $\alpha$
values of -1,1 and 0 is proportional to $x,y$ and $z$, respectively.

It is convenient to decompose the sum over $\beta$ on the right hand
side of Eq. \ref{y9} for the $\alpha^{th}$ component as
\begin{eqnarray}
 M_\alpha \equiv <\tilde{\psi}_f|p_\alpha |\tilde{\psi}_i>& = &\sum_{\gamma} M^\gamma_\alpha\label{y12}\\
 &=& \sum_{\gamma} \sum_{\delta}M^{\gamma,\delta}_\alpha\label{y13}\\
 &=&\sum_{\gamma} \sum_{\delta}\sum_{L,L'}M^{\gamma,\delta}_{\alpha,L,L'}
 \label{y14}
 \end{eqnarray}
with 
\begin{equation}
M^{\gamma,\delta}_{\alpha,L,L'} =  i^{l-l'-1} ({C_{L'}^{\gamma,\delta}})^* 
C_L^{\gamma,\delta}  B_{l,l'}^{\gamma,\delta} {\cal G}_{L,L'}^\alpha
\label{y15}
\end{equation}
The indices $\gamma$ and $\delta$ in Eqs. \ref{y12}-\ref{y14} are
keyed in with the pristine (tetragonal) crystal structure of BISCO
detailed in Section IIIA below, and taken together, encompass the sum
over all atoms $\beta$ in the unit cell in Eq. \ref{y8}.  After the
sum over the angular momentum channels $L$ and $L'$ in Eq. \ref{y14}
has been carried out, the index $\delta$ which only takes two values
(1 and 2) in Eq. \ref{y13} sums over the contributions of pairs of
atoms (other than the $Ca$ atom) related by mirror symmetry of the
lattice with respect to the $Ca$ layer. In this sense, $\delta$ may be
thought of as a "pairing" index. $\gamma$ then takes on eight distinct
values which include seven pairs of atoms $(Bi, O_{Bi}, Sr, O_{Sr},
Cu, O_{Cu,x}, O_{Cu,y})$ and the $Ca$ atom as a "site index". The
aforementioned indexing is by no means unique and other useful
variations can be envisioned in the present case and certainly for
other lattice types.

\section{Relevant features of the lattice geometry and electronic 
  structure of BISCO}

\subsection{Lattice structure}

Since various structural features of the lattice are invoked
frequently in the discussion, it is appropriate that we state our
nomenclature clearly. For this purpose, two different cross-sections
through the conventional body centered tetragonal unit cell, together
with the detailed arrangement of atoms in different layers, are shown
in Fig. 1.  The distinct atom "pairs" to go with the index $\gamma$ in
Eqs. \ref{y12}-\ref{y14} are identified in Fig. 1(c). Values of
$\gamma$ from 2-8 involve an atom and its mirror partner reflected in
the $Ca$ layer, yielding a total of 15 basis atoms per primitive
lattice point or 30 atoms in the conventional unit cell.  The $Ca$
atom ($\gamma=1$) is seen to lie at the center of symmetry in the
upper part of Fig. 1(a). The two $CuO_2$ planes sit above and below
the $Ca$ layer and consist of the two $Cu$ atoms in the basis set
denoted by the ($\gamma, \delta$) indices (2,1) and (2,2). $O_{Cu,x}$
and $O_{Cu,y}$ are given similarly by $\gamma=3$ and 4, each with
$\delta=1,2$; note that although $O_{Cu,x}$ and $O_{Cu,y}$ are
identical in terms of the crystal potential in the tetragonal case, it
is still useful to make this distinction because states along $x$ and
$y$ will in general respond differently depending upon the direction
of polarization of the incident light. Next come the $SrO$ layers with
$Sr$ ($\gamma=5$) and the associated apical $O_{Sr}$ atoms
($\gamma=6$), followed by the $BiO$ layers involving the $Bi$
($\gamma=7$) and $O_{Bi}$ ($\gamma=8$) atoms.

\subsection{Electronic structure}

Fig. 2 shows the energy bands in BISCO along the symmetry lines
$\Gamma-M-X-\Gamma$. The crystal potential underlying these
computations differs somewhat from the selfconsistent LDA potential in
that the $Bi$-$O$ pockets around the M-point have been lifted above
the $E_F$ in order to account for their absence in the ARPES
measurements. The two bands in the vicinity of the $E_F$ are related
to the $CuO_2$ planes; the splitting between $a$ and $b$ is the
bilayer splitting arising from interaction between the two $CuO_2$
planes. At the M-point, the wavefunction of the upper band $a$ is
antibonding or antisymmetric with respect to the $Ca$ plane (along the
$z$ direction) while that of the lower band $b$ is bonding or
symmetric. The size of the bilayer splitting is generally larger along
$\Gamma-M$ compared to the $\Gamma-X$ direction. We have chosen to
work in the present simulations with a potential that yields a rather
large bilayer splitting because this makes it easier to identify the
signatures of the bonding and antibonding bands $b$ and $a$ in the
theoretical ARPES spectra\cite{foot7}. However, computations have been
repeated for numerous other cases where the bilayer splitting was
varied by placing suitable potential barriers between the $CuO_2$
layers, and the results indicate that our conclusions concerning the
nature of these states and their behavior in the ARPES spectra are
robust to such variations.

The analysis of the recent high resolution ARPES spectra leaves little
doubt that the Fermi surface of BISCO consists of two distinct
sheets\cite{matri,mlcape,maria}.  One is the standard large hole-like
sheet centered around the $X$ or $Y$ symmetry point, and a second
sheet arising from a band lying very close the the $E_F$ which could
be either slightly hole- or electron-like. The precise position of the
$E_F$ will depend of course on the doping level of the system. In any
event, it is often useful to avoid convoluting the computed spectra
with the Fermi function so that the nature of the spectra over a wider
range of ${\bf k_\parallel}$ and energy values, including unoccupied
states above the $E_F$, can be elucidated. Notably, the ${\bf
  k_\perp}$ dispersion of bands terminating at points $a$ and $b$,
even though it varies with ${\bf k_\parallel}$, is quite small (order
of a few meV at $\overline{M}$) and is not important in much of our
discussion.

\section{Salient features of the ARPES spectra from the $\overline{M}$ point}

\subsection{Photon energy dependence of emission intensity from 
bonding and antibonding states}

Fig. 3 shows the ARPES intensity from the antibonding and bonding
states ($a$ and $b$) at the $\overline{M}$ point as a function of the
photon energy. For our illustrative purposes we have taken the light
to be polarized such that only the $x$-component of the vector
potential ${\bf A}$ is non-zero. The initial state width corresponds
to $\Sigma_i''=100$ meV \cite{foot4}.  The final state width is set
using $\Sigma_f''=1$ eV$-$ we return to this point in Section IVB
below. The emission intensity is seen to change dramatically with
photon energy. The bonding state emits strongly around 18 and 23 eV
with a feature around 33 eV, while the antibonding state possesses a
somewhat broad region of high emissions extending from 14-26 eV. The
intensity for both states is relatively small from 27-40 eV and below
11 eV. These results have significant implications for ARPES
experiments. For example, emission maps from the $E_F$ can be expected
to look quite different at various photon energies; at an energy such
as 22 eV, the emissions from the pristine lattice will be dominant,
while at an energy such as 28 eV, the contributions due to modulations
of the underlying symmetry (e.g. orthorhombic distortion, superlattice
modulation, etc.)  will generally become more prominent\cite{maria}.
Energies such as 18 or 23 eV will tend to highlight the bonding state
while the antibonding state will likely be more intense around 22 eV,
effects of final state broadening and other factors notwithstanding.
We should particularly keep in mind that there are inherent
uncertainities of a few eV's in locating the final states in the first
principles computations, and that the positions of various features in
Fig. 3 will generally possess similar uncertainty.  We have carried
out additional computations (not shown in the interest of brevity) and
find that the aforementioned type of variations in intensity which
selectively highlight different aspects of the electronic structure
continue to be manifest at photon energies much higher than the 40 eV
upper limit in Fig. 3.

\subsection{Effect of final state width}

Fig. 4 shows how the plot of Fig. 3 for the antibonding initial state
changes when the final state width varies corresponding to
$\Sigma_f''$ values from 20 meV to 2 eV\cite{foot4}.  The results for
small $\Sigma_f''$ make it clear that the spectrum is intrinsically
made up of numerous transitions with different weights. With
increasing $\Sigma_f''$ these transitions broaden and overlap to
produce varying spectral shapes of Fig. 4. For $\Sigma_f'' < 100$ meV,
little change in shape occurs because the spectra are controlled by
the initial and not the final state width, the former having a
constant value given by $\Sigma_i''=100$ meV in all cases.  Note that
to reproduce the observed linewidths, $\Sigma_f''$ used in the
computations will need to account not only for the damping of the
final state reflected via the imaginary part of the self-energy, but
also the effects of experimental parameters such as the temperature of
the measurement and the acceptor angle of the analyzer (typically of
order of a few tenths of a degree in the high resolution ARPES
setups). In this sense, $\Sigma_f''$ is perhaps better viewed as a
computational rather than a physical parameter in the theory with
values ranging from 1-4 eV, where 4 eV is appropriate for 40-60 eV
photons \cite{strocov96,courths99}.  Incidentally, computations become
more demanding as the value of $\Sigma_f''$ decreases because the mean
free path of the outgoing electron increases and an exponentially
larger number of layers must be included in the calculations to obtain
a converged result.

\subsection{Polarization dependence}

Insofar as the polarization dependence is concerned, note first that
in Fig. 5 light is assumed incident normally to the surface and that
the detector lies in the $x-z$ plane with the $x$-axis defined by the
$Cu-O_x$ bond direction; the polarization vector of light is varied
holding all other experimental parameters fixed. The inserts in the
shape of number eight are polar plots of the intensities of various
peaks ($a$, $b$, $c$ and $c'$) as a function of the azimuthal angle
$\phi$ between the polarization vector and the $x$-axis. The intensity
is maximum for all peaks at $\phi=0$, i.e. when the light is polarized
along the $Cu-O_x$ bonds. Under a $90^\circ$ rotation the intensity
nearly vanishes as the polarization vector turns perpendicular to the
plane of the detector.

Insight into the polarization dependence discussed in the preceding
paragraph can be gained by examining the behavior of Eq. \ref{y5} for
the geometrical setup of Fig. 5. Since the detector lies in the $x-z$
plane of mirror symmetry, the final state $|f>$ must obviously possess
even symmetry with respect to this plane in order to be observable. On
the other hand, $p_x$ or $p_z$ (or, equivalently, $\partial V/\partial
x$ or $\partial V/\partial y$ from Eq. \ref {y7}) is even but $p_y$ is
odd under reflections in this mirror plane.  It follows then that only
the matrix elements $<f|p_x|i>$ and $<f|p_z|i>$ can be non-zero for
{\it even} initial states while for {\it odd} initial states only
$<f|p_y|i>$ can be non-zero. We now express A in spherical polar
coordinates as
\begin{equation}
{\bf A} =  A_0{\bf \hat{e}} = A_0( \cos \phi \cos \theta {\bf \hat{e}}_x + 
\sin \phi \cos \theta {\bf
\hat{e}}_y + \sin \theta {\bf \hat{e}}_z)
\label{y16}
\end{equation}
where $A_0$ is the amplitude, $\theta$ and $\phi$ are the polar and
azimuthal angles of the incoming light beam, and ${\bf
  \hat{e}}_\alpha$ are unit vectors along the Cartesian axis. For the
setup of Fig. 5, $\theta=0$, and Eq. \ref{y5} reduces to
\begin{equation}
I =   {e\hbar}/{mc}\sum_i B_{ii} | A_0 (\cos \phi {\bf \hat{e}}_x +
\sin \phi  {\bf \hat{e}}_y) \cdot <\psi_f|({\bf p}_x+{\bf p}_y)|\psi_i>|^2
\label{y17}
\end{equation}
It turns out that the antibonding state $a$ as well as the bonding
state $b$ is even and, therefore, recalling the aforementioned
symmetry arguments, only the $p_x$ term contributes and Eq. \ref{y17}
yields
\begin{equation}
I =  {e\hbar}/{mc} \sum_i B_{ii} | A_0  <\psi_f|{p}_x|\psi_i>|^2 \cos ^2 \phi
\label{y18}
\end{equation}
which explains the shape of polar plots of Fig. 5 for states $a$ and
$b$; states $c$ and $c'$ must also be even in view of the shapes of
the related polar plots. The preceding analysis is generalized
straightforwardly to consider vector potentials with other non-zero
components. If a non-zero $p_z$ component is included, Eq. \ref{y18}
is modified with a radially symmetric contribution.  In this vein, if
the mirror symmetry plane is rotated by $45^\circ$ as is the case in
YBCO (neglecting the effect of Cu-O chains), that leads to polar plots
which are rotated similarly.\cite{santa,polar}

Fig. 5 also shows the energy distribution curve (EDC) for 22 eV
photons when ${\bf k_\parallel}$ is held fixed at $\overline{M}$.  The
EDC for the case when the initial state width is taken to be a small
constant is given by the thin line for reference in order to reveal
the spectral peaks clearly. More realistically, the initial state
width increases with increasing binding energy.  This effect is
modelled by the EDC of the thick line which shows that much of the
structure for binding energies greater than $\sim$ 100 meV is washed
out by lifetime effects.

\subsection{Final state band structure, momentum matrix element}

Fig. 6 compares the one-step ARPES intensity at the $\overline{M}$
point for the antibonding state $a$ with the corresponding result
based on the bulk momentum matrix element of Eq. \ref{y12} and helps
address a number of related issues. The small final state width used
in these simulations allows the intrinsic spectral structure to be
seen clearly. The final state bands shown as a function of $k_\perp$
in the lower portion of the figure provide a map of all available
transitions since $k_\perp$ is not conserved in the photoemission
process. [Note that the horizontal scale has been converted to refer
to the photon energy.]  It is striking that although these bands
appear to form a semi-continuum, there are gaps in the underlying
final state spectrum and the situation thus is quite unlike the free
electron case\cite{foot3}. Also, $k_\perp$ dispersion is not always
negligible with some bands dispersing as much as 0.5 eV despite the
small size of the BZ in the z-direction.

As already noted, it tends to be difficult to associate peaks in the
one-step spectra with transitions to specific final states. The
problem becomes exacerbated when states are broadened to incorporate
finite lifetime effects. Considerable insight into the behavior of the
ARPES matrix element can nevertheless be gained via the momentum
matrix element $<\tilde{\psi}_f|{\bf p}|\tilde{\psi}_i>$ of Eq.
\ref{y9} for bulk transitions. We emphasize however that
$<\tilde{\psi}_f|{\bf p}|\tilde{\psi}_i>$ does not properly account
for the processes of transport and ejection of the photoelectron as is
the case in the one-step ARPES computation. Moreover, since
$<\tilde{\psi}_f|{\bf p}|\tilde{\psi}_i>$ involves the bulk
wavefuctions $\tilde{\psi}_i$ and $\tilde{\psi}_f$, modifications of
these wavefunctions in the vicinity of the surface and the effect of
possible formation of surface states are not incorporated. Even so,
Fig. 6 shows that the one-step spectrum for the antibonding state $a$
is rather similar to that based on the momentum matrix element. Some
differences in the intensities of certain peaks in the figure may be
understood as consequences of the fact that the momentum matrix
element has been evaluated at $k_\perp=0$ while in the one-step result
$k_\perp$ value is in general non-zero and varies from $k_\perp=0$ to
the zone boundaries at $k_\perp=\pm\pi/c$ for various peaks.  Results
for the bonding state $b$ (not shown for brevity) are similar in that
the level of agreement between the one-step and the momentum matrix
element based spectra is comparable to that of Fig. 6.

\section{Nature of transition matrix elements and origin of spectral 
         features in terms of character of initial and final states}

\subsection{Contributions of different atomic sites and 
  related interferences}

Contributions $M^{\gamma,\delta}$ of different atomic sites in the
unit cell can be delineated via the decomposition of Eqs.
\ref{y12}-\ref{y14}. Table 1 and Fig. 7 provide illustrative examples
of such an analysis. Considering $x$-polarized light, values of
$M_x^{\gamma,\delta}$ for each of the 15 basis atoms for one
particular transition from the states $a$ and $b$ at M are listed in
Table 1. Fig. 7 on the other hand gives various significant
contributions to {\it all} transitions from state $a$ over the energy
range of 10-27 eV.

The strong influence of the mirror symmetry around the $Ca$ layers is
evident in Table 1. For the particular transition considered, the
magnitudes of $M_x^{\gamma,\delta}$ are the same for the two mirror
partners corresponding to the two $\delta$ values, but the phases
differ by $\pi$. In fact, contributions cancel pairwise for the
bonding state $b$, but interfere constructively for the antibonding
state $a$. For many other transitions from $a$ and $b$ the situation
is quite the opposite in that this sort of interference is {\it
  destructive} for $a$, but {\it constructive} for $b$.  As we move
away from M, these cancellation effects become less perfect as the
phases do not always differ exactly by $\pi$ and/or the magnitudes of
the two mirror partners are no longer the same.

Figure 7 provides a somewhat broader perspective. Total matrix element
$|M_x|$ obtained by summing over $\gamma$ and $\delta$ is depicted
pictorially for many transitions from $a$ together with contributions
from pairs of $Cu$, $O_{Cu,x}$, $O_{Bi}$ and $O_{Sr}$ atoms;
$M^{\gamma}_x$ for $Ca$, $Sr$, $Bi$ and $O_{Cu,y}$ are less than 0.01
in magnitude and are not shown. Of a total of 64 possible transitions
from state $a$ over the energy range of Fig. 7, only 20 transitions
that are shown possess a significant intensity. A few of the
transitions are identified by numbers 1-7 for convenient reference. As
already noted above in connection with Table 1, transition 1 is quite
intense and arises almost completely from $O_{Cu,x}$. Transition 4
contains an admixture of $O_{Cu,x}$ and $Cu$, while transition 3
contains additionally a contribution from $O_{Bi}$. Transition 5
contains contribution from all these atoms as well as $O_{Sr}$, but as
indicated by downarrows, $Cu$ and $O_{Bi}$ give negative contributions
yielding a net reduction in the total intensity. These examples would
make it clear that constructive or destructive interference can take
place not only for a given atomic pair, but also between different
atomic sites. We emphasize that the intensity variations in Fig. 7
reflect changes in the character of the final states since the initial
state for all transitions is fixed to be the same.

\subsection{Contributions of various angular momentum channels and 
  interference effects therein}

Further insight can be gained by considering the angular momentum
dependence of the matrix element, i.e. the quantities
$M^{\gamma\delta}_{x,L,L'}$ of Eq. \ref{y15}.  For illustrative
purposes, we consider first the aforementioned transition 1 with the
help of Tables II-V, followed by a few comments concerning the
behavior of transitions more generally via Fig. 8.

The weights in different angular momentum channels (defined as
$\sum\limits_\delta |C_L^{\gamma\delta}|^2$) for the initial states
$a$ and $b$ associated with various atomic sites are given in Tables
II and III, while the final state for transition 1 is characterized
similarly in Table IV\cite{footnote15}.  We emphasize that these
weights depend upon details of how space within the unit cell is
ascribed to different atoms and the size and shape of the interstitial
region. For this reason, we should keep in mind that the results of
Tables II-IV possess inherent uncertainty, even though these are
representative of the character of the states involved. As expected,
wavefunctions of both $a$ and $b$ are seen from Tables II and III to
be predominantly $Cu(d_{x^2-y^2})$ and $O_{Cu,x}(p_x)$ like; $O_{Bi}$
as well as $O_{Sr}$ states hybridize more strongly with the bonding
level $b$ compared to $a$.  The final state in Table IV continues to
be weighted in favor of $Cu$ and $O_{Cu,x}$, but being 17.7 eV above
the initial states, there is substantial change with respect to
distribution among angular momentum channels and atomic sites.

We turn next to Table V and the matrix element for transition 1 from
state $a$. Of the large number of possible terms ($\sim 300$) under
the summations in Eq. \ref{y14}, only the few elements
$M^{\gamma}_{x,L,L'}$ listed in Table V possess substantial magnitude.
The remaining elements are either zero due to selection rules or are
negligibly small. The $p_x\rightarrow d_{x^2-y^2}$ channel on
$O_{Cu,x}$ is seen to be the strongest with a contribution (in
arbitrary units) of 0.27. It is most striking that the net
contribution from $Cu$ is only 0.02 even though the initial state is
dominated by $Cu$. This is the result of strong interference effects:
On $Cu$, different angular momentum channels interfere destructively
as manifest in the positive and negative signs of elements, while on
$O_{Cu,x}$, there is a constructive interference yielding a large
total intensity of 0.43.  These observations hint that the ARPES
matrix element can display remarkable site selectivity properties. For
transition 1 under discussion, for example, almost all the
photoelectrons emanate from $O_{Cu,x}$ sites with little contribution
from other sites in the crystal due to complicated interferences.

Fig. 8 considers all transitions from $a$ over the energy range of
10-27 eV.  The total matrix element $|M_x|$ and the portion
$|M_x^\gamma|$ associated with $O_{Cu,x}$ in the top two panels of the
figure are reproduced from Fig. 7 for ease of discussion; the total
$O_{Cu,x}$ intensity is decomposed into angular momentum channels in
the three lower panels. By comparing the lengths of arrows in Figs.
8(a) and 8(b), we see that for most transitions (but not all)
$O_{Cu,x}$ component is strong. The effect however is not as dominant
as in the case of transition 1; for example, only about 60 percent of
transitions 2-4 come from $O_{Cu,x}$ and transition 6 has essentially
a zero $O_{Cu,x}$ part.  Moreover, even though different angular
momentum channels interfere constructively for most transitions in
that arrows are mostly pointed up in Figs. 8(c)-(e), this is not
always the case as some arrows are pointed downwards. In particular,
for transition 5 roughly half of the intensity in $p_x\rightarrow s$
channel is destroyed by inteference from $p_x\rightarrow d_{x^2-y^2}$
channel, and for transition 7, these two channels destroy each other
nearly completely.  We note finally that, of the many factors involved
in defining $M^{\gamma\delta}_{\alpha,L,L'}$ on the right side of Eq.
17, much of the energy dependence depicted in Fig. 8 originates from
the behavior of the wavefunction coefficients of the final states
($C_{L'}^{\gamma\delta}$); the Gaunt coefficients ${\cal
  G}_{L,L'}^\alpha$ are constants and the quantities
$B_{l,l'}^{\gamma,\delta}$ involving integrals of radial wavefunctions
are slowly varying functions of energy.

\subsection{$k_\perp$ dependence}

So far we have focused on the excitation of states $a$ and $b$ at the
M($\pi$,0,0) point, i.e. for $k_\perp=0$. Changes in the momentum
matrix element as a function of $k_\perp$ for several selected
transitions from the initial state $a$ are considered in Fig. 9.
Transitions 1-4 have been discussed previously in connection with
Figs. 7 and 8. Transitions 2b and 3b however, which lie quite close in
energy to transitions 2 and 3, respectively, were not shown before
since their intensity is zero or near-zero at $k_\perp=0$, but as Fig.
9 shows, this is not the case for $k_\perp\ne 0$. Transitions 1 and 4
decrease slightly in intensity as $k_\perp$ increases from zero to
$\pi /c$. On the other hand, transitions 2, 2b, 3 or 3b undergo large
changes. Observable effects after spectra are smoothed with realistic
broadenings of the order of a few eV's will likely be much smaller
since the intensities of transitions 2 and 2b (or similarly 3 and 3b)
taken together are expected to change little with $k_\perp$.

We have carried out additional computations at ${\bf k}$-points on the
$\overline{\Gamma}-\overline{M}$ line away from $\overline{M}$.  The
results continue to display complicated interference effects involving
pairs of atoms and/or various atomic sites, essentially along the
lines of the preceding commentary concerning emissions from
$\overline{M}$, and require little further elaboration.

\section{Summary and conclusions}

We have investigated the nature of the ARPES spectrum from BISCO using
the first-principles band theory framework. Our focus is on
understanding emission from electronic states along the
$\overline{\Gamma}-\overline{M}$ symmetry line and particularly from
the bonding and antibonding combinations of bands formed from the
states associated with the $CuO_2$ bilayers which have been the
subject of considerable recent discussion and controversy. Salient
features of the ARPES spectrum delineated are as follows.

1. The intensity of the antibonding as well as the bonding state at
$\overline{M}$ displays a strong dependence on photon energy even
though the two states differ significantly from each other in this
respect. This result indicates that the bonding state will be
highlighted at certain photon energies while the antibonding state
will be intense at other energies; moreover, at photon energies where
emission from either state is relatively weak, we would expect the
effects of deviations of the underlying lattice from perfect
tetragonal symmetry (i.e. via various modulations) to become more
visible.

2. The spectra depend strongly on the polarization of the incident
light.  At the $\overline{M}$ point, for states in the vicinity of the
$E_F$, the intensity is maximum when light is polarized along the $x$-
or the $Cu-O_x$ bond direction, but becomes nearly zero upon a
$90^\circ$ rotation of the polarization vector perpendicular to the
plane of the detector. Simple arguments are presented to explain the
$cos^2\phi$ (where $\phi$ is the angle between the $x$-axis and the
polarization vector) variation observed in the one-step computations
which yield a polar intensity plot in the shape of the number eight in
BISCO and other cuprates.

3. An examination of the final states indicates that, despite the
appearance of a semi-continuum, there are gaps in the final state
spectrum and that the $k_\perp$ dispersion of some bands is as large
as 0.5 eV even though the Brillouin zone size is small along the
z-direction. The situation thus is quite unlike the free electron
case.

4. We show how considerable insight into the nature of the {\it
  specific} ARPES excitations can be adduced in terms of the momentum
matrix element for {\it bulk} transitions within the solid in
combination with constraints appropriate for a particular ARPES
measurement. Such transition specific information tends to be
difficult to ascertain within the more comprehensive one-step
framework, especially in the presence of finite initial and final
state dampings.

5. With the preceding point in mind, we have studied the nature of the
momentum matrix element for emission from the $CuO_2$ plane related
antibonding state $a$ and the bonding state $b$ at $\overline{M}$ at
length over the transition energy range of 10-27 eV. We show how the
matrix element, $|M|^2\equiv |<\tilde{\psi}_i|{\bf
  p}|\tilde{\psi}_f>|^2$, can be decomposed usefully into
contributions from various atomic "pairs" in the unit cell (i.e. an
atom and its mirror partner obtained by reflection through the $Ca$
layer about which the BISCO lattice is symmetric), and further into
different angular momentum channels.  Contributions from the two atoms
in the aforementioned atomic pairs display remarkable effects of
interfering {\it constructively} for some transitions and {\it
  destructively} for others. The net contribution from the $O_{Cu,x}$
pairs seems to dominate the spectrum in many cases even though the
initial and final states involved possess large weights on the Cu
sites. These strong interference effects continue to be present in
different angular momentum channels. The analysis indicates that many
transitions are controlled essentially by just the {\it single} $p_x
\rightarrow d_{x^2-y^2}$ channel on the $O_{Cu,x}$ sites. This
striking result hints that the ARPES matrix element may possess most
interesting site selectivity properties.

6. The intensity of many transitions increases (or decreases)
substantially as a function of $k_\perp$. However, these variations
are usually coupled with nearby transitions whose intensity decreases
(or increases) so that the net contribution is generally not expected
to show a strong $k_\perp$ dependence.

In summary, the present analysis makes it clear that the complicated
nature of the ARPES matrix element in complex materials hides
considerable potential for extracting useful information via this
spectroscopy about the nature of specific states. Studies of other
cuprates along these lines should allow an identification of robust
aspects of the electronic structure related to the $CuO_2$ planes.

\acknowledgements

It is a pleasure to acknowledge important conversations with Bob
Markiewicz. This work is supported by the US Department of Energy
contract W-31-109-ENG-38, and benefited from the allocation of
supercomputer time at the NERSC and the Northeastern University
Advanced Scientific Computation Center (ASCC) and Institute of
Advanced Computing (IAC), Tampere.  One of us (S.S.) acknowledges
Suomen Akatemia, Vilho, Yrj\"o ja Kalle V\"ais\"al\"an Rahasto,
Tekniikan Edist\"amiss\"a\"ati\"o and Jenny ja Antti Wihurin Rahasto
for financial support.

\begin{table}%[h]
\caption{
Magnitude and phase of the matrix element $M^{\gamma,\delta}_x$ is given for
various sites $\gamma$ and $\delta$ for transition 1 in Fig. 7 (or Fig. 8)
for both the antibonding initial state $a$ and the bonding initial state $b$
at M. Note that Figs. 7 and 8 only refer to state $a$, but that this table
also gives results for the state $b$. } \vspace*{0.5cm}
\begin{tabular}{|l|l|l|d|d|d|d|}
element&$\gamma$&$\delta$& \multicolumn{2}{c|}{Antibonding state $a$}&\multicolumn{2}{c|}{Bonding state $b$}\\
  \tableline
&&&$\mid M^{\gamma,\delta}_x\mid$&$phase$&$\mid M^{\gamma,\delta}_x\mid$&$phase$ \\ 
  \tableline
 Ca &1&1&0.0279&35.6&0.0000&  -\\
 Cu &2&1&0.0209&35.6&0.0191&  16.8\\
 Cu &2&2&0.0209&35.6&0.0191&-163.2\\
 O$_{Cu,x}$&3&1&0.4309&35.6&0.3111&  16.8\\
 O$_{Cu,x}$&3&2&0.4309&35.6&0.3111& -163.2\\
 O$_{Cu,y}$&4&1&0.0518&35.6&0.0784&   16.8\\
 O$_{Cu,y}$&4&2&0.0518&35.6&0.0784& -163.2\\
 Sr &5&1&0.0158&35.6&0.0081&  16.8\\
 Sr &5&2&0.0158&35.6&0.0081&-163.2\\
 O$_{Sr}$ &6&1&0.0316&35.6&0.0140  & 16.8\\ 
 O$_{Sr}$ &6&2&0.0316&35.6&0.0140 &-163.2\\ 
 Bi &7&1&0.0314&-144.4&0.0366&-163.2\\
 Bi &7&2&0.0314&-144.4&0.0366&  16.8\\
 O$_{Bi}$ &8&1&0.0167&-144.4& 0.0417& -163.2\\
 O$_{Bi}$ &8&2&0.0167&-144.4& 0.0417 & 16.8\\
\end{tabular}
\end{table} 
\narrowtext

\begin{table}%[h]

\caption{ 
Wavefunction of the antibonding state $a$ at M-point is characterized in terms of 
weights (in percent) defined as $\sum\limits_{\delta} |C_L^{\gamma\delta}|^2$ 
associated with different sites $\gamma$ (see Eq. \ref{y8}). Not all angular 
momentum channels are listed. Channels with small weights are omitted for 
brevity; the total weight for this reason adds 
to less than 100 percent.\protect\cite{footnote15} }

\vspace*{0.5cm}

\begin{tabular}{|l|r|r|r|r|r|r|}
$\gamma$         & Tot. & $s$ & $p_z$ & $p_x$ & $d_{3z^2-r^2}$ & $d_{x^2-y^2}$ \\ 
  \tableline
 Ca &&&&&& \\
  Bi          & 12   &          & 7 & 5    && \\
   Sr &&&&&& \\
    Cu          & 43   & 3 &           &          & 12 & 28  \\
     O$_{Cu,x}$ & 24   &          &           & 24  && \\
      O$_{Cu,y}$ &&&&&& \\
       O$_{Bi}$  & 4    &          & 4  &&& \\
        O$_{Sr}$  & 6    &          & 6 &&&\\ 
\end{tabular}
\end{table} 
 \narrowtext

\begin{table}%[h]

\caption{ Same as the caption to Table II, except that this table refers to 
the bonding state $b$ at the M-point. }

\vspace*{0.5cm}

\begin{tabular}{|l|r|r|r|r|r|r| }
$\gamma$ & Tot. & $s$ & $p_z$ & $p_x$ & $d_{3z^2-r^2}$ & $d_{x^2-y^2}$ \\ 
   \tableline
Ca  &&&&&& \\
Bi          & 10   & 4  & 6  &&& \\
Sr &&&&&&\\
Cu          & 29   & 2  &          &          & 7  & 20  \\
O$_{Cu,x}$ & 13   &          &          & 13 && \\
O$_{Cu,y}$ & 4    &          & 4  &&& \\
O$_{Bi}$  & 24   &          & 6 &18  && \\
O$_{Sr}$  & 11   &          & 11 &&& \\
\end{tabular}
\end{table} 
 \narrowtext

\begin{table}%[h]

\caption{ Same as the caption to Table II, except that this table refers to 
the final state of transition 1 in Fig. 7 (or Fig. 8) involving the 
excitation of the antibonding initial state $a$. } 

\vspace*{0.5cm}
\begin{tabular}{|l|r|r|r|r|r|r|r|}
$\gamma$  & Tot. & $s$   & $p_z$   & $p_x$    & $d_{xz}$  &$d_{x^2-y^2}$& $f_{z^3}$\\ 
\tableline
Ca          & 7   &             & 2  &              &             &             &5\\
Bi          &      &             &             &              &             &             &  \\
Sr          & 18   & 10 & 4 &              &             &             &4 \\
Cu          & 29   &             &             & 29&             &             & \\
O$_{Cu,x}$  & 18   & 9 & 3 &              &             & 6& \\
O$_{Cu,y}$  & 4    &             &             &  2 & 2&             & \\
O$_{Bi}$    &      &             &&&&& \\
O$_{Sr}$    &      &             &&&&& \\ 
\end{tabular}
\end{table} 
 \narrowtext

\begin{table}%[h]

\caption{
Components $\sum\limits_\delta M_{x,L,L'}^{\gamma\delta}$ 
of the momentum matrix element 
(see Eqs. 14-17) for transition 1 
marked in Fig. 7 (or Fig. 8) which involves excitation
of the antibonding initial state $a$ at M-point. Only sites and channels
with significant weight are listed. }

\vspace*{0.5cm}

\begin{tabular}{|c|l|d|d|}
$\gamma$      & Channel: $L \rightarrow L'$ &$M_{x,L,L'}^{\gamma}$&  $\sum\limits_{LL'}M_{x,L,L'}^{\gamma}$\\  
  \tableline
O$_{Cu,x}$ & $p_x \rightarrow s$                       & +0.12      & \\
& $p_x \rightarrow d_{3z^2-r^2}$            & +0.04      & \\
& $p_x \rightarrow d_{x^2-y^2}$             & +0.27      & +0.43 \\ 
  \tableline
Cu         & $s \rightarrow p_x$                       & +0.01    & \\
& $d_{x^2-y^2} \rightarrow p_x$             & +0.07      & \\
& $d_{x^2-y^2} \rightarrow f_{xz^2}$        & -0.02      & \\
& $d_{x^2-y^2} \rightarrow f_{x(x^2-3y^2)}$ & -0.03      & \\
& $d_{3z^2-r^2} \rightarrow p_x$              & +0.03    & \\
& $d_{3z^2-r^2} \rightarrow f_{xz^2}$         & -0.04    & +0.02\\
\end{tabular}

\end{table}

\begin{figure}
\begin{center}
%\hbox{\epsfig{file=fig1.eps,width=8.5cm,angle=00} }                  
\end{center}             
\caption{
  Body-centered tetragonal crystal structure of Bi2212.  (a) and (b)
  give (110) and (100) cross-sections through the structure,
  respectively. (c) depicts the arrangement of atoms in different
  layers.  Various basis atoms are identified by the pairs of indices
  ($\gamma,\delta$) discussed in the text.  }
\label{fig1}               
\end{figure}      

\begin{figure}
    \begin{center}
%    \hbox{\epsfig{file=fig2.eps,width=8.5cm,angle=00} }
       \end{center}
\caption{
  Band structure of Bi2212 in the vicinity of the Fermi energy $E_F$
  (shown by the horizontal dashed line) along the symmetry lines
  $\Gamma-M-X-\Gamma$.  Note that the $Bi$-$O$ pockets around M in the
  standard LDA band structure have been lifted above the $E_F$; see
  text for details of crystal potential. $a$ and $b$ are the
  antibonding and bonding states at M, respectively, arising from the
  $CuO_2$ planes. $c$ and $c'$ are $Bi$-$O$ related states.  }
\label{fig2}
\end{figure}

\begin{figure}
    \begin{center}
%       \hbox{\epsfig{file=fig3.eps,width=8.5cm,angle=00} }
              \end{center}
      \caption{
        ARPES intensity for emission from the $CuO_2$ plane related
        antibonding and bonding states at $\overline{M}$ as a function
        of the photon energy.  }
\label{fig3}
\end{figure}

\begin{figure}
\begin{center}
%\hbox{\epsfig{file=fig4.eps,width=8.5cm,angle=90} }
\end{center}
\caption{
  ARPES intensity from the antibonding state $a$ at $\overline{M}$ as
  a function of the photon energy for four different values of the
  width of the final state determined by the parameter $\Sigma_f''$
  \protect\cite{foot4}.  }
\label{fig4}
\end{figure}

\begin{figure}
    \begin{center}
%       \hbox{\epsfig{file=fig5.eps,width=8.5cm,angle=90} }
              \end{center}
      \caption{
        Polarization dependence of the ARPES spectra at $\overline{M}$
        for 22 eV light. Energy distribution curve (EDC) including
        realistic initial state broadening for light polarized along
        the $x$-axis is shown (thick line); thin line gives the EDC
        computed by using an artificially small damping of initial
        states to reveal the spectral peaks $a$, $b$, $c$ and $c'$
        clearly.  EDC's have not been convoluted with the Fermi
        function.  Inserts in the shapes of number eight on top of
        various peaks are polar plots of peak intensity as the
        direction of the polarization vector ($\phi$) is varied and
        possess the $cos^2\phi$ dependence in accord with Eq.
        \ref{y18}. $\phi$ is measured with respect to the $x$-axis or
        the $Cu-O_x$ bond direction; as seen in the inset in the top
        part of the figure, the detector (D) is assumed to lie in the
        vertical plane through the $x$-axis. }
\label{fig5}
\end{figure}

\begin{figure}
\begin{center}
%\hbox{\epsfig{file=fig6.eps,width=8.5cm,angle=00} }
\end{center}
\caption{
  Computed one-step ARPES intensity in BISCO at $\overline{M}$ from
  the antibonding state $a$ as a function of the photon energy is
  compared with the corresponding result based on the momentum matrix
  element $|M_x|^2$ for bulk transitions (see Eq. \ref{y12}). A small
  final state damping is used to reveal features clearly. The relevant
  final state bands are shown as a function of $k_\perp$ and energy in
  the lower figure where the energy scale has been shifted to
  correspond to photon energy.  }

\label{fig6}
\end{figure}

\begin{figure}
\begin{center}
%\hbox{\epsfig{file=fig7.eps,width=8.5cm,angle=00} }
\end{center}
\caption{
  Total matrix element $|M_x|$ (topmost panel) for various transitions
  from the antibonding state $a$ over the transition energy range of
  10-27 eV is decomposed into contributions from various "pairs" of
  atomic sites $\gamma$ (panels (b)-(e)) as described in Eqs.
  \ref{y12}-\ref{y14}. Down arrows indicate a negative value of
  $M_x^\gamma$. Sites not shown yield a negligible contribution.
  Transitions labelled 1-7 (connected by dashed lines for ease of
  identification across the panels) are discussed in the text to
  highlight various interference effects.  }
\label{fig7}
\end{figure}

\begin{figure}
\begin{center}
%\hbox{\epsfig{file=fig8.eps,width=8.5cm,angle=00} }
\end{center}
\caption{ 
  Same as the caption to Figure 7 with the two top panels (a) and (b)
  reproduced from Figure 7 for ease of reference. Panels (c)-(e) give
  the decomposition of $O_{Cu,x}$ results in (b) into various angular
  momentum channels $\sum\limits_\delta M_{x,L,L'}^{\gamma\delta}$
  (see Eq. \ref{y14}). }
\label{fig8}
\end{figure}

\begin{figure}
\begin{center}
%\hbox{\epsfig{file=fig9.eps,width=8.5cm,angle=00} }
\end{center}
\caption{
  $k_\perp$ dependence of the total momentum matrix element $|M_x|$
  for a few selected transitions from the antibonding initial state
  $a$ at $\overline{M}$.  Transitions 1-4 are the same as those in
  Figs. 7 and 8. Transition 2b (or 3b) possesses energy close to 2 (or
  3) but its intensity is zero (or quite small) for $k_\perp=0$, or
  equivalently, at $\overline{M}(\pi,0,0)$.  }
\label{fig9}
\end{figure}


\begin{thebibliography}{99}
  
\bibitem{matri} A. Bansil and M. Lindroos, Phys. Rev. Lett.  {\bf 83},
  5154(1999).

\bibitem{mlcape}A. Bansil and M. Lindroos, J. Phys. Chem. Solids {\bf
    59}, 1879(1998).

\bibitem{maria}M. C. Asensio, J. Avila, L. Roca, A. Tejeda, G.D. Gu,
  M. Lindroos, R.S. Markiewicz, and A. Bansil, to be published.

\bibitem {zx2}P.V. Bogdanov, et al., Phys. Rev. Lett. (2001),
  cond-mat/0005394.
  
\bibitem {zx3}D.L. Feng, et al., Phys. Rev. Lett. {\bf 86},
  5550(2001).
  
\bibitem {des2}Y.D. Chuang, et al., Phys. Rev. Lett. {\bf 83},
  3717(1999); Y.D. Chuang, et al., cond-mat/0107002.

\bibitem{gene1}See e.g., ``Spectroscopies in Novel Superconductors'',
  edited by A. Bansil, R. Markiewicz, S. Sridhar and D. Liebenberg [J.
  Phys. Chem. Solids {\bf 59}, No 10-12 (1998)], and other volumes in
  this series for discussions of these and related issues.

\bibitem{pendry2} J.B. Pendry, Surface Sci. {\bf 57}, 679(1976);
  J.F.L. Hopkinson, J.B. Pendry and D.J. Titterington, Computer Phys.
  Commun.  {\bf 19}, 69(1981).

\bibitem{stanf}A. Bansil and M. Lindroos, J. Phys. Chem. Solids {\bf
    56}, 1855(1995).

\bibitem{ncco}M. Lindroos and A. Bansil, Phys. Rev. Lett. {\bf 75},
  1182(1995).

\bibitem{cucis} M. Lindroos and A. Bansil, Phys. Rev. Lett. {\bf 77},
  2985(1996).

\bibitem{larsson} C.G. Larsson, Surface Science {\bf 152/153},
  213(1985).

\bibitem{cardona78} M. Cardona and L. Ley, {\em Photoemission in
    Solids I: General Principles}, Spinger-Verlag Berlin 1978.
  
\bibitem{kunz79} C. Kunz in {\em Photoemission in Solids II}, edited
  by L. Ley and M. Cardona, Springer-Verlag Berlin 1979, p313.
  
\bibitem{inglesfield92} J.E. Inglesfield and E.W. Plummer on S.D.
  Kevan (ed.), {\em Angle Resolved Photoemission, Theory and Current
    Applications}, Elsevier 1992.
  
\bibitem{lindroos82} M. Lindroos, Physica Scripta {\bf 25}, 788(1982).
  
\bibitem{ni100}S. Sahrakorpi, M. Lindroos and A. Bansil, submitted to
  Phys. Rev. B.
  
\bibitem{santa} A. Bansil et al., J. Phys. Chem. Solids {\bf 54},
  1185(1993); M. Lindroos et al., Physica {\bf C212}, 347(1993).

\bibitem{abri} K. Gofron, J. C. Campuzano, A. A. Abrikosov, M.
  Lindroos, A. Bansil, H. Ding, D. Koelling and B. Dabrowski, Phys. Rev. Lett. {\bf
    73}, 3302(1994).
  
\bibitem{photogen}Angle-resolved Photoemission, Theory and Current
  Applications edited by S.D. Kevan, Elsevier, 1992.
  
\bibitem{mahan}G.D. Mahan, Phys. Rev. {\bf B2}, 4334(1970).
  
\bibitem{caroli} C. Caroli, D. Lederer-Rozenblatt, B. Roulet, D.
  Saint-James, Phys. Rev. {\bf B8}, 4552(1973).
  
\bibitem{durham}P.J. Durham, B.L. Gyorffy and A.J. Pindor, J. Phys. F:
  Metal Phys. {\bf 10}, 661(1980).
  
\bibitem{braun}J. Braun, G. Th\"orner and G. Borstel, Phys. Stat. Sol.
  {\bf B130}, 643(1985).
  
\bibitem{grass} M. Grass, J. Braun and G. Borstel, Phys. Rev. {\bf
    B50}, 14827(1994).
  
\bibitem {camp2}H.M. Fretwell, et al., Phys. Rev. Lett. 84,
  4449(2000).
  
\bibitem {gold1}S.V. Borisenko, et al., Phys. Rev. Lett. {\bf 84},
  4453(2000); A.A. Kordyuk, et al., cond-mat/0104294.
  
\bibitem{takahashi}T. Sato, et al., Phys. Rev. B{\bf 63},
  132502(2001).
  
\bibitem{aebi}P. Aebi, et al., Phys. Rev. Lett. {\bf 72}, 2757(1994).
  
\bibitem {camp1}C.F. Behrens, et al., Phys. Rev. Lett. {\bf 76},
  1553(1996).
  
\bibitem{bianconi}N. Saini, et al., Phys. Rev. Lett. {\bf 79},
  3467(1997).
  
\bibitem{shen}X.J. Zhou, et al., Science {\bf 286}, 268(1999).
  
\bibitem{valla}T. Valla, et al., Phys. Rev. Lett. {\bf 85}, 828(2000).
  
\bibitem{lindau}A. Zakharov et al. , Phys. Rev. B {\bf 61}, 115(2000).
  
\bibitem{onellion}S. Rast et al., Europhys. Lett.  {\bf 51},
  103(2000).
  
\bibitem{footmatti} Wave length of light is $\sim 600$ \AA\ while the
  unit cell dimensions are $\sim 30$ \AA.
  
\bibitem{seppo} Seppo Sahrakorpi, Ph.D. Thesis, Tampere University of
  Technology (2001), to be published.
  
\bibitem{segall}B. Segall, Phys. Rev. 105, 108(1957).
  
\bibitem{peter} P. E. Mijnarends and A. Bansil, J. Phys.:Condens.
  Matter 2, 911(1990).
  
\bibitem{foot7} It is straightforward to mimic correlation effects in
  the present framework by introducing a self-energy correction to the
  Hamiltonian. Further work is however needed in this regard,
  especially for incorporating strong correlations effects in the
  ARPES simulations.
  
\bibitem{foot4} $\Sigma_i''$ and $\Sigma_f''$ are the imaginary parts
  of the initial and final state self-energies, respectively, used to
  include lifetime effects.  For a damped propagator the
  full-width-at-half-maximum (FWHM) of the peak is roughly given by $2
  \Sigma''$. See, N.V. Smith, P.Thiry and Y. Petroff, Phys. Rev. B
  {\bf 47}, 15476(1993).
  
\bibitem{strocov96} V.N. Strocov, H.I. Starnberg and P.O. Nilsson, J.
  Phys.: Condens. Matter {\bf 8}, 7539 (1996).
  
\bibitem{courths99} R. Courths, S. L\"obus, S. Halilov, T.
  Scheunemann, H. Gollisch and R. Feder, Phys. Rev. B {\bf 60}, 8055
  (1999).
  
\bibitem{polar} M. Lindroos and A. Bansil, J. Phys. Chem. Solids {\bf
    52}, 1447(1991).
  
\bibitem{foot3}When the states are broadened, the effect of these gaps
  will of course be reduced.
  
\bibitem{footnote15} While these weights are relevant for analyzing
  the behavior of the momentum matrix element, one should keep in mind
  that these do not include the effect of the radial part of the
  wavefunction as seen from Eqs. 10 and 11.

\end{thebibliography}
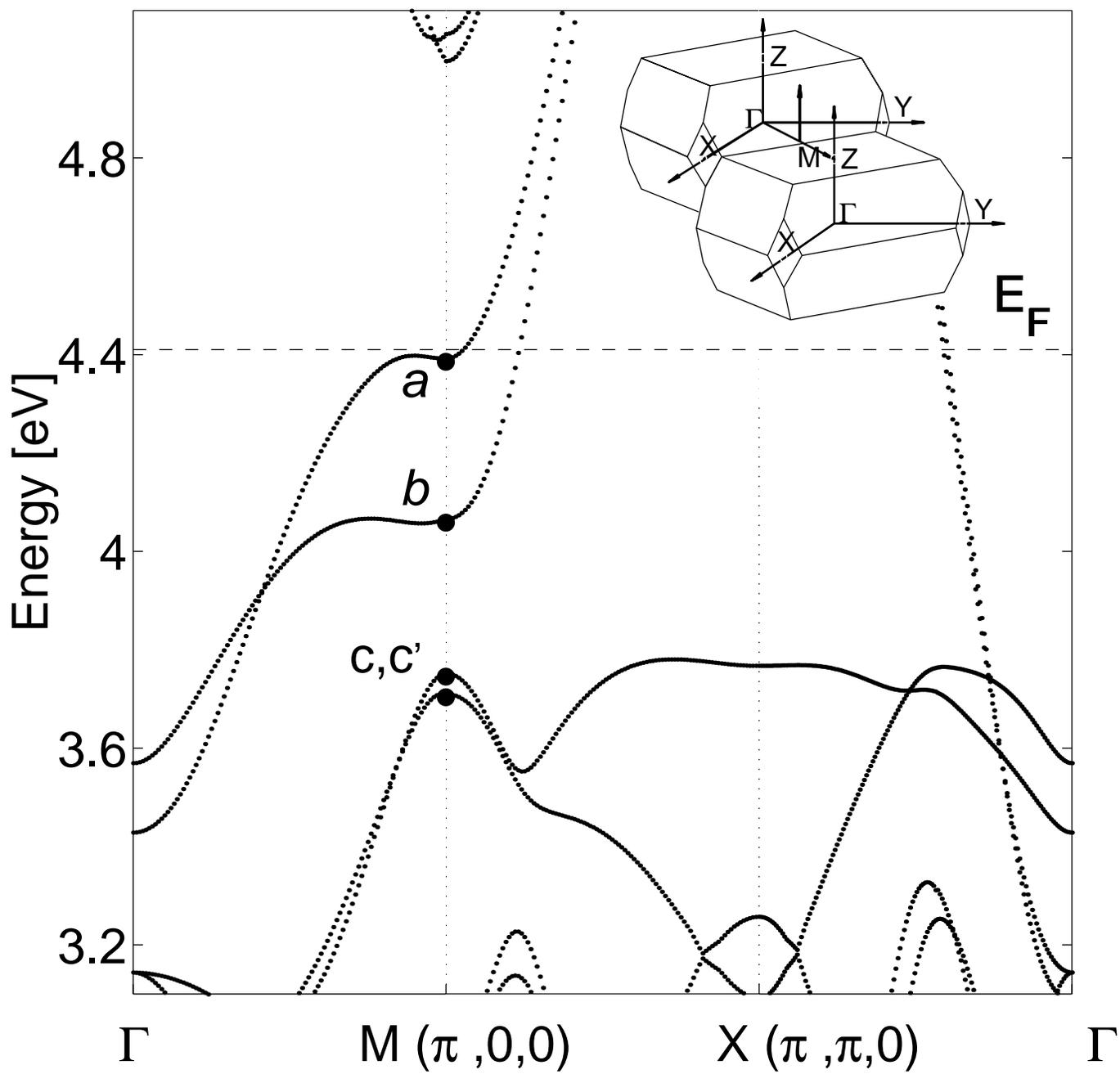
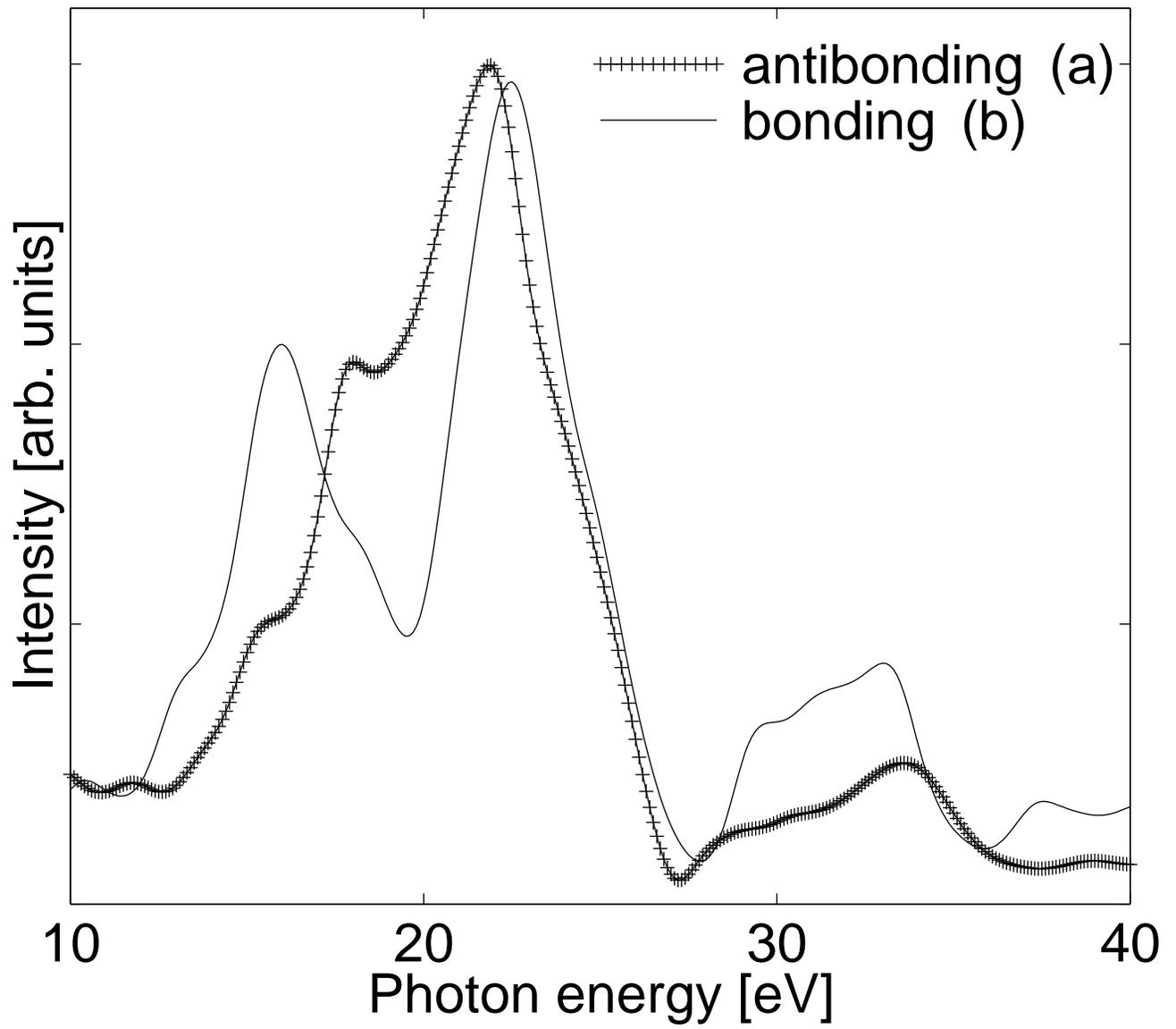
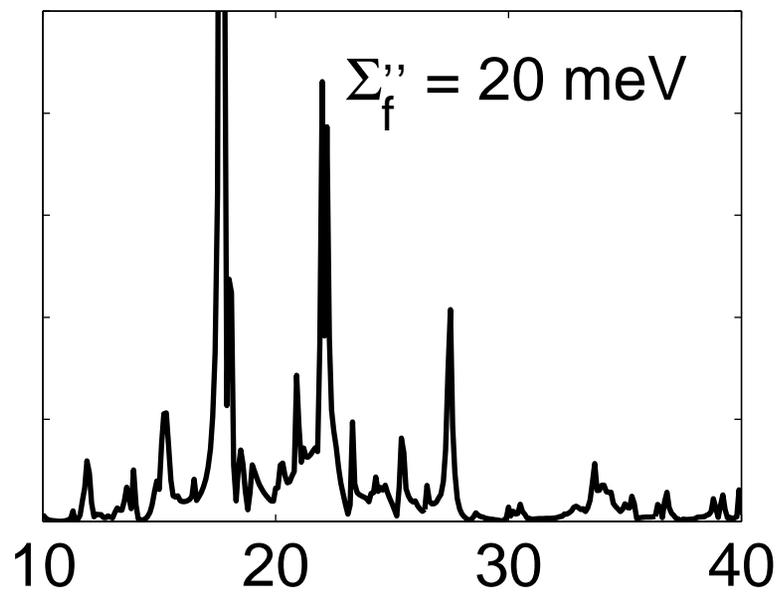
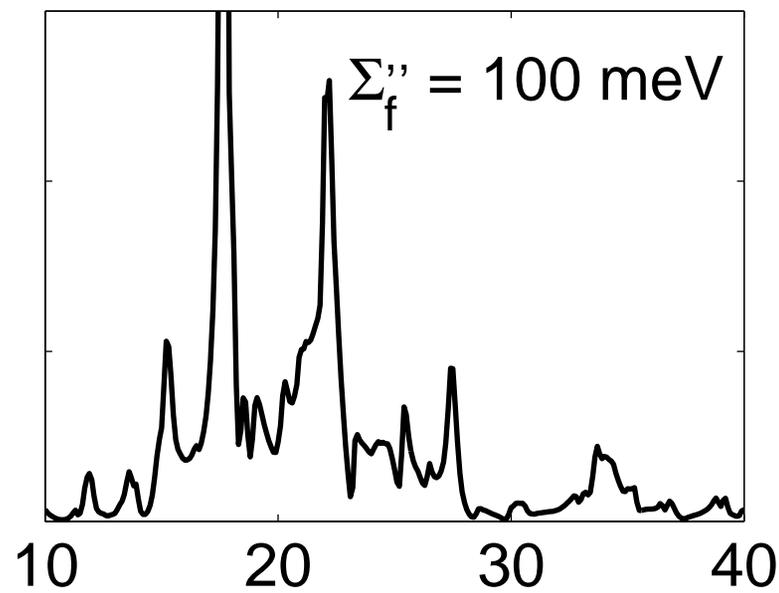
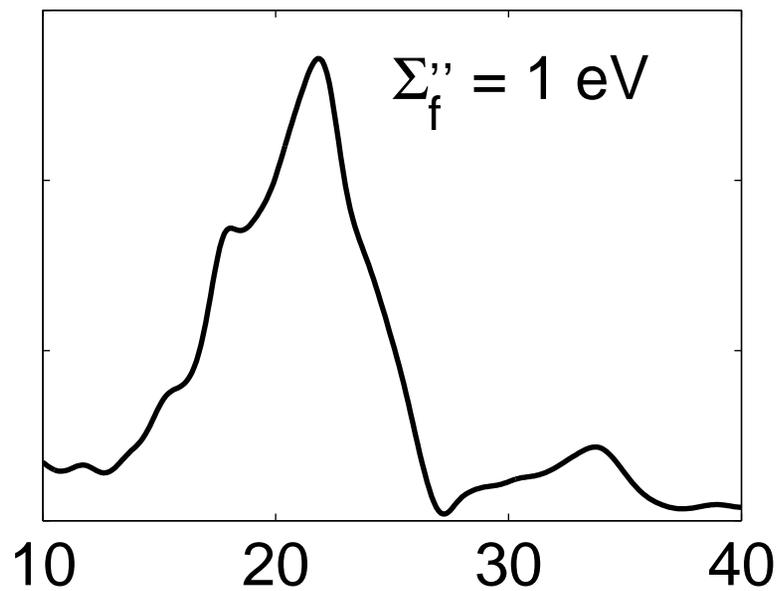
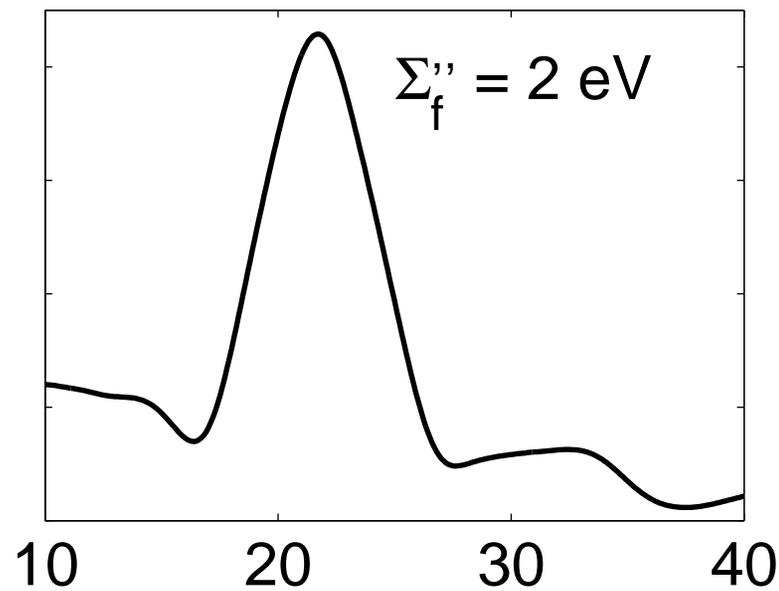
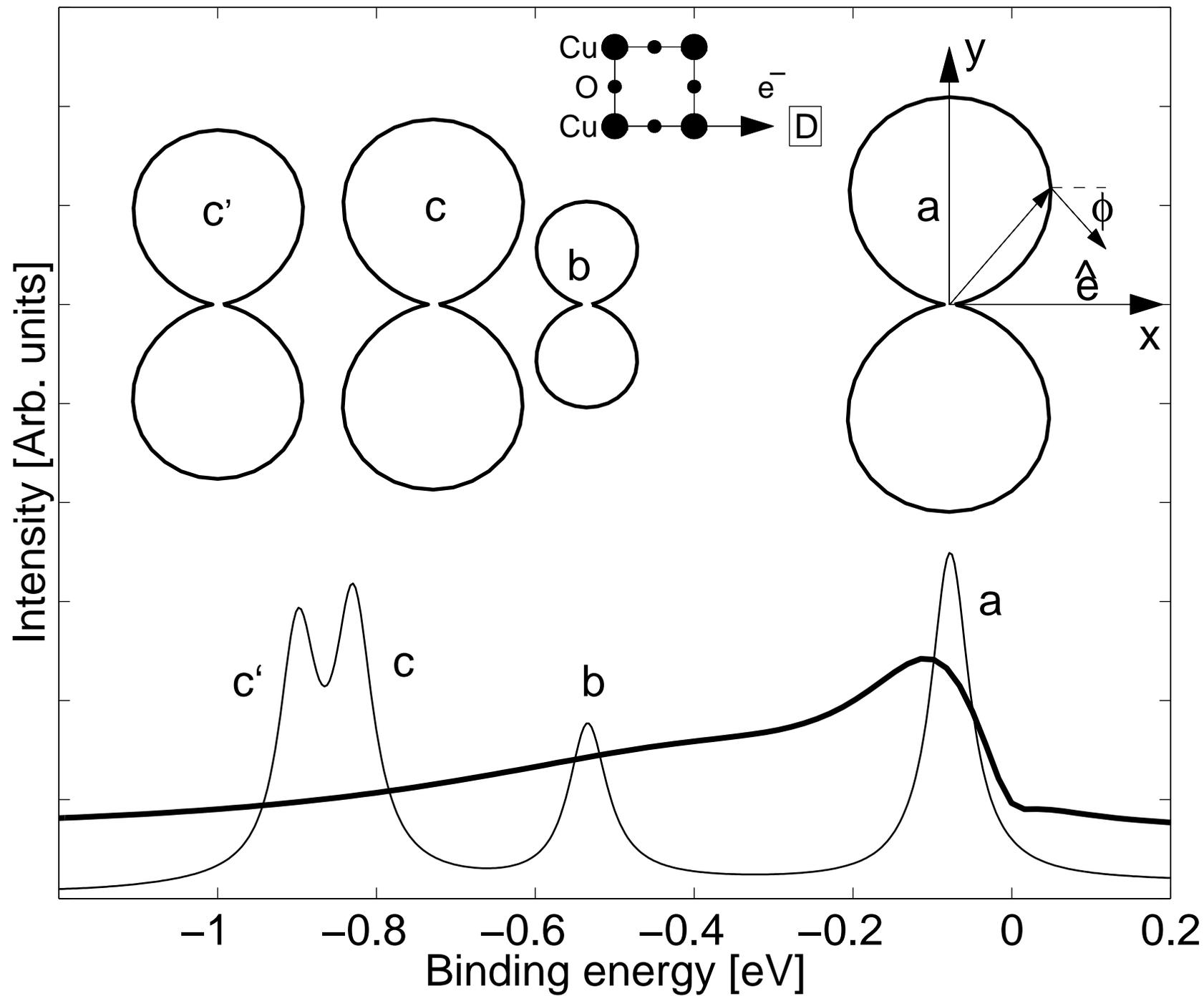
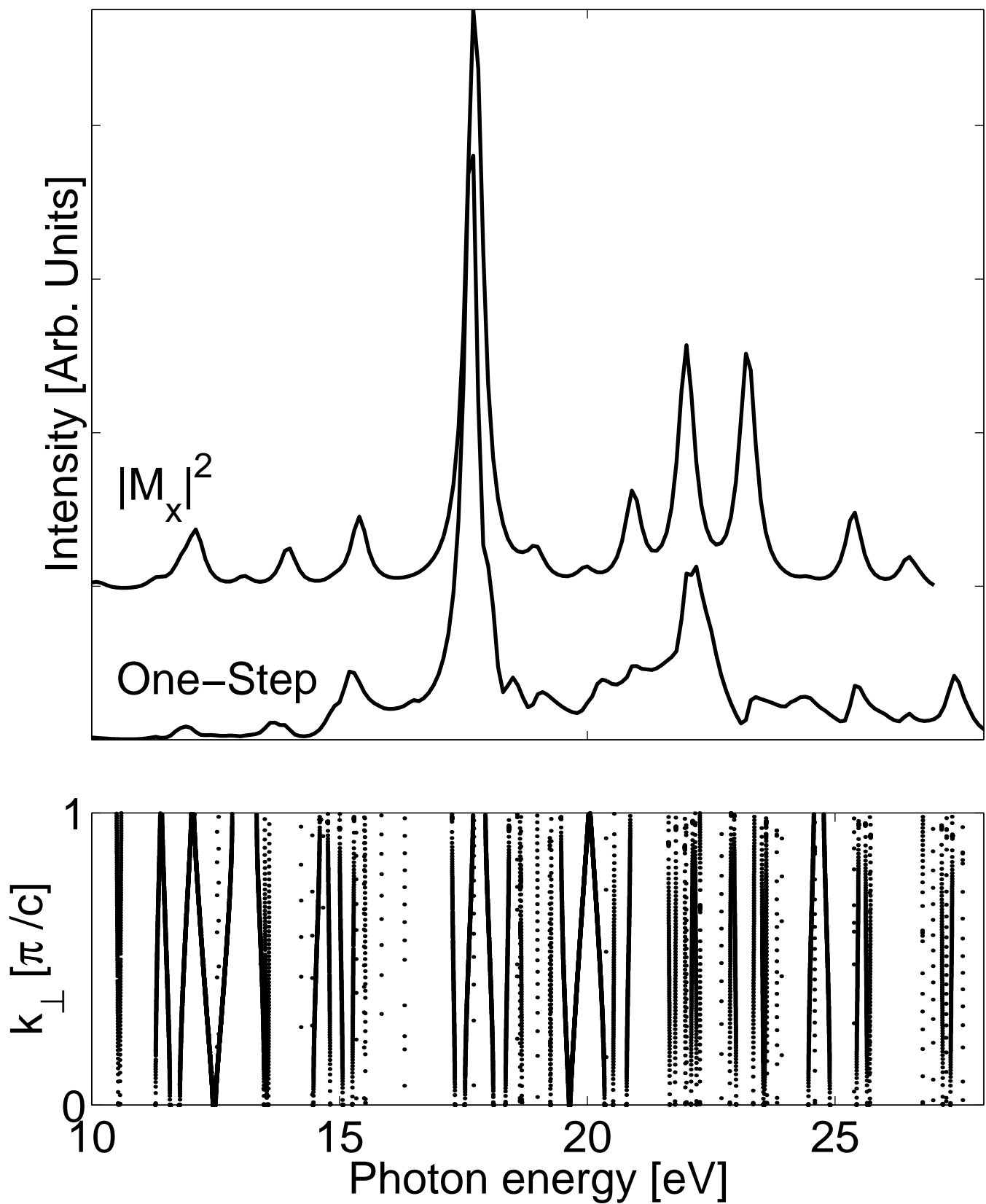
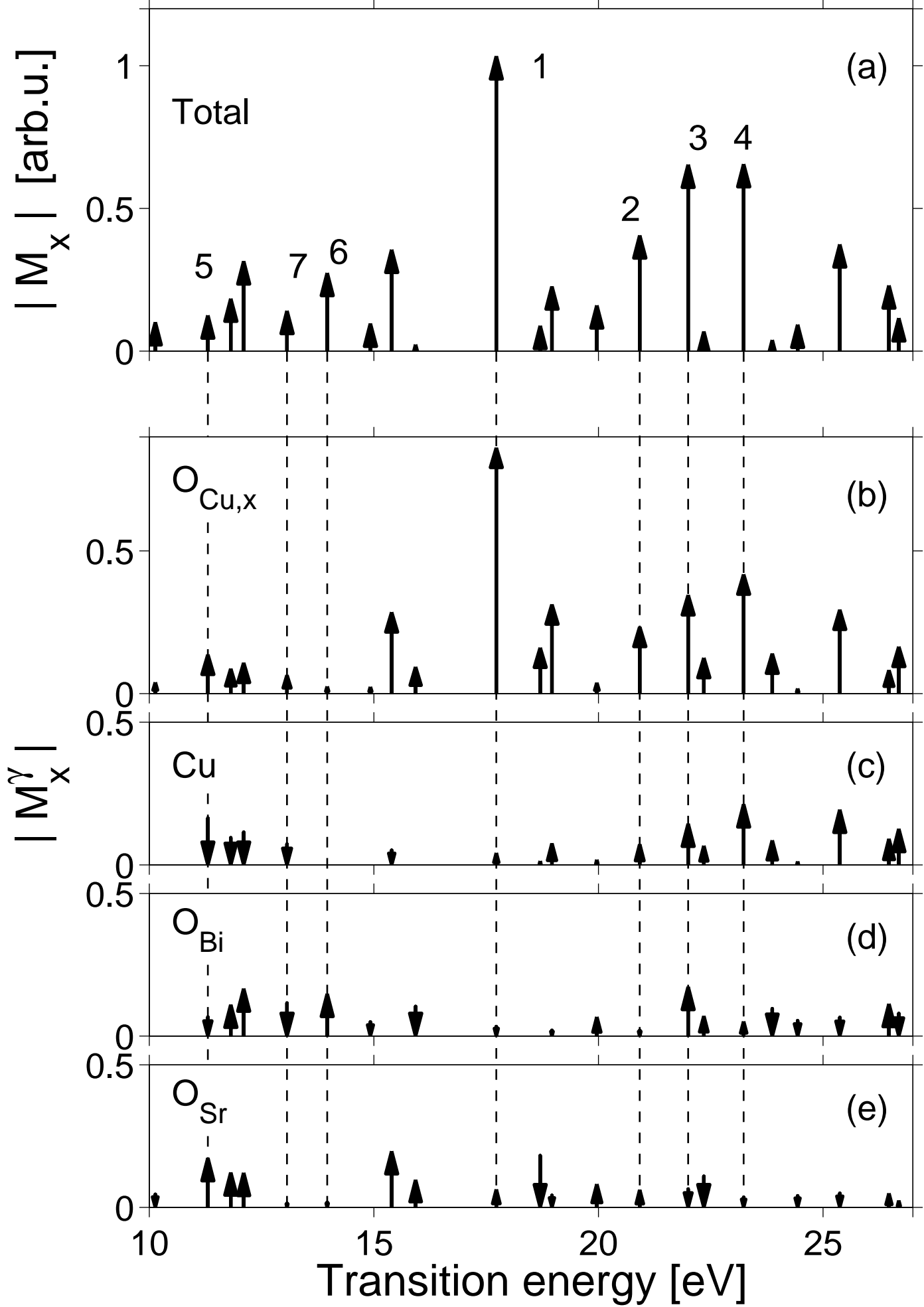
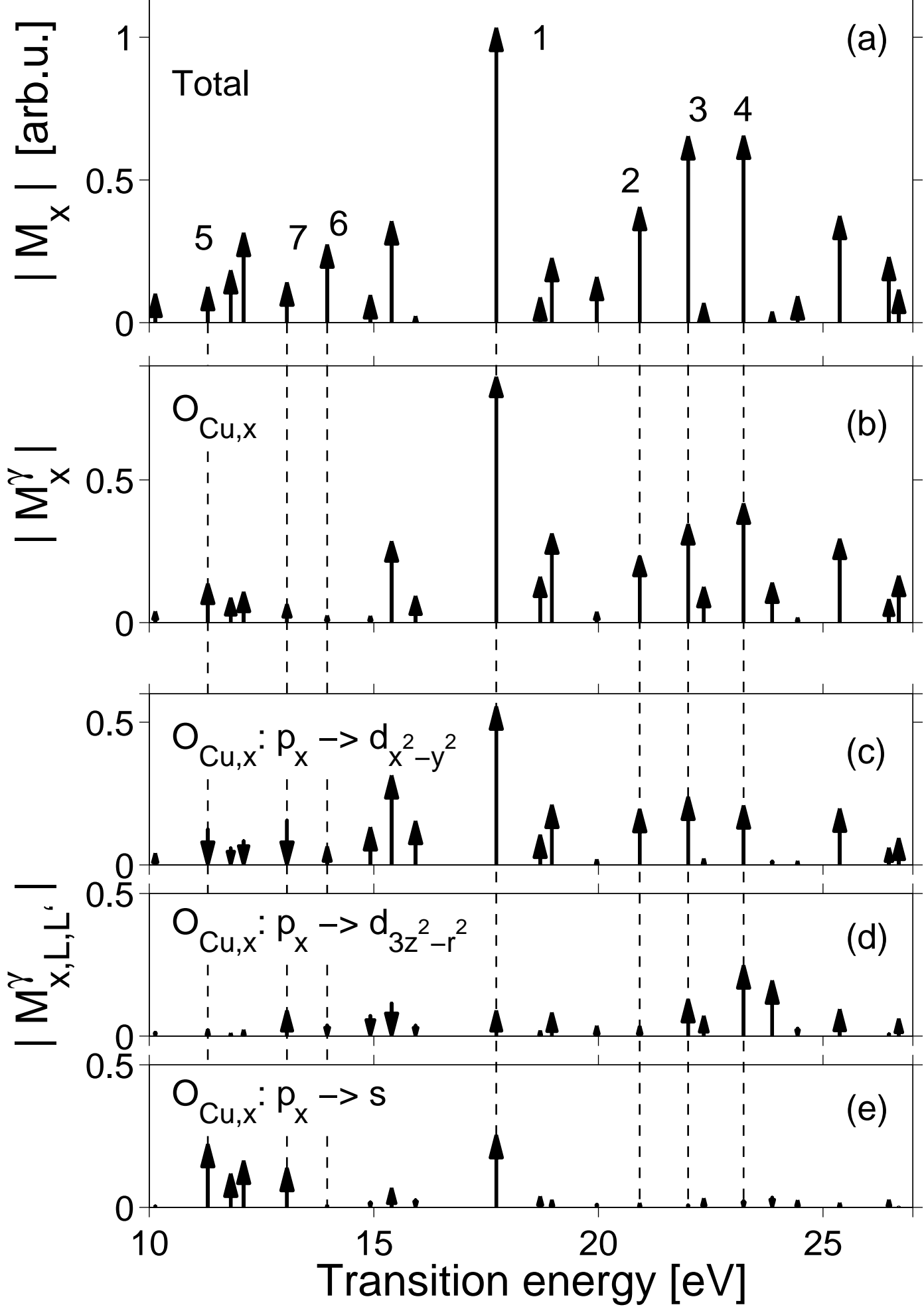
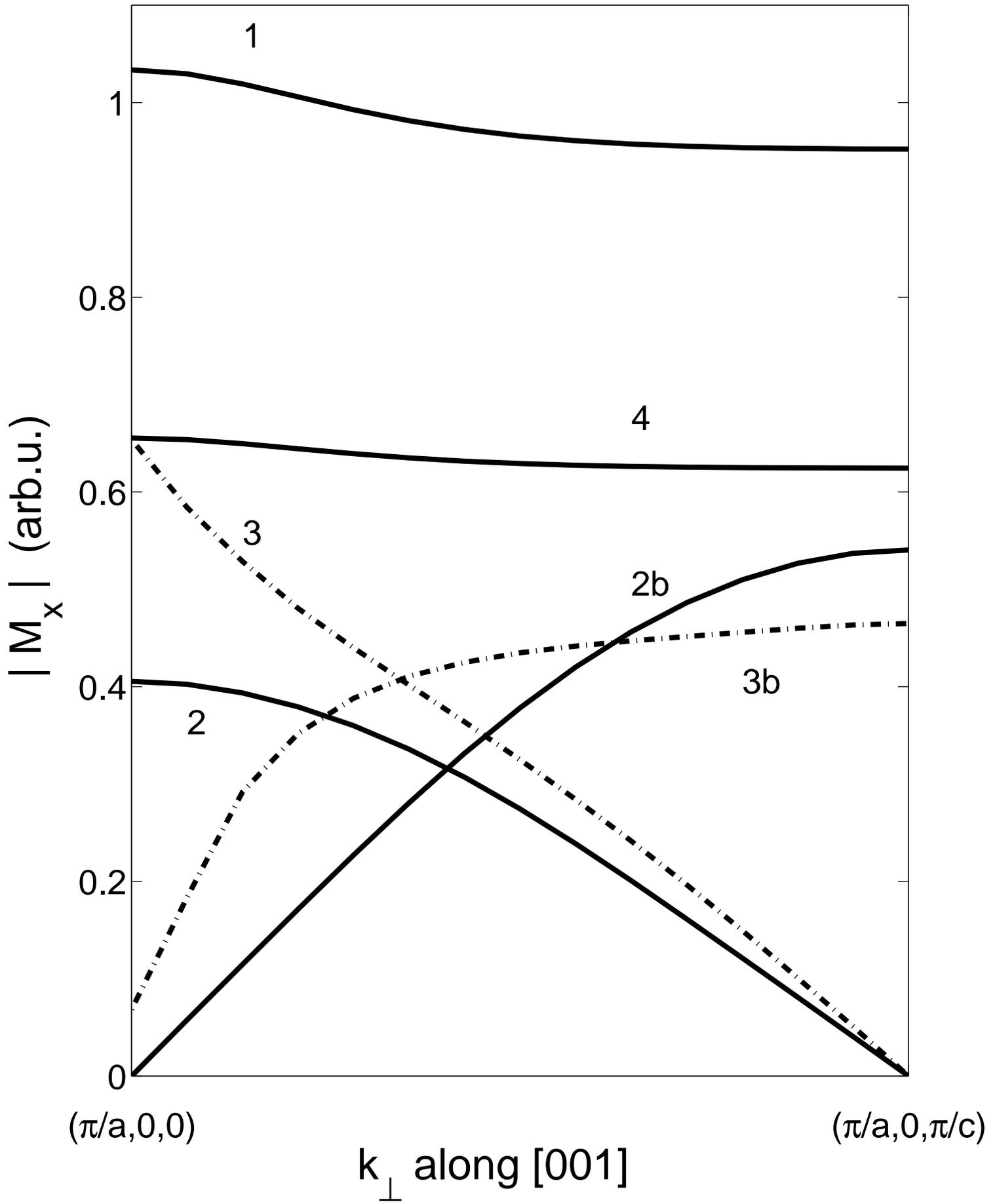
\end{document}